\documentclass[12pt]{article}
\usepackage{latexsym}
\usepackage{amsmath}
\usepackage{amssymb}
\usepackage{epsfig,graphics}
\usepackage{graphicx}
\usepackage{booktabs}
\usepackage{multirow}
\usepackage{tikz}
\usepackage{bm}
\usepackage{graphicx}
\usepackage{epsf}
\usepackage{epsfig}
\usepackage{color}
\usepackage{dcolumn}     
\usepackage{amsfonts,amssymb}
\usepackage{dsfont}
\usepackage{slashed}
\usepackage{subfigure}

\newcommand{\be}{\begin{equation}}
\newcommand{\ee}{\end{equation}}

\newcommand{\bea}{\begin{eqnarray}}
\newcommand{\eea}{\end{eqnarray}}

\begin{document}

\begin{titlepage}

\vspace*{0.6in}
 
\begin{center}
{\large\bf On the mass of the world-sheet `axion' \\
in $SU(N)$ gauge theories in $3+1$ dimensions.}\\
\vspace*{0.75in}
{Andreas Athenodorou$^{a,b}$ and Michael Teper$^{c}$\\
\vspace*{.25in}
$^{a}$Computation-based Science and Technology Research Center, The Cyprus Institute, 20 Kavafi Str., Nicosia 2121, Cyprus \\
\vspace*{.1in}
$^{b}$Department of Physics, University of Cyprus, POB 20537, 1678 Nicosia, Cyprus\\
\vspace*{.1in}
$^{c}$Rudolf Peierls Centre for Theoretical Physics, University of Oxford,\\
1 Keble Road, Oxford OX1 3NP, UK}
\end{center}

\vspace*{0.4in}

\begin{center}
{\bf Abstract}
\end{center}

There is numerical evidence that the world sheet action of the confining flux tube
in $D=3+1$  $SU(N)$ gauge theories contains a massive excitation with $0^-$ quantum
numbers whose mass shows some decrease as one goes from $SU(3)$ to $SU(5)$. It has
furthermore been shown that this particle is naturally described as arising from
a topological interaction term in the world-sheet action,
so that one can describe it as being `axion'-like.
Recently it has been pointed out that if the mass of this `axion'  vanishes
as $N\to\infty$ then it becomes possible for the world sheet theory
to be integrable in the planar limit. In this paper we perform lattice calculations of
this `axion' mass from $SU(2)$ to $SU(12)$, which allows us to make a controlled
extrapolation to $N=\infty$ and so test this interesting possibility.
We find that the `axion'  does not in fact
become massless as $N\to\infty$. So if the theory is to possess planar
integrability then it must be some other world sheet excitation that becomes
massless in the planar limit.

\vspace*{0.95in}

\leftline{{\it E-mail:} a.athenodorou@cyi.ac.cy, mike.teper@physics.ox.ac.uk}

\end{titlepage}

\setcounter{page}{1}
\newpage
\pagestyle{plain}

\tableofcontents

\section{Introduction}
\label{section_intro}

The spectrum and world sheet action of confining flux tubes in $D=3+1$ $SU(N)$ gauge
theories is now known to be, for the most part, remarkably simple. That is to say,
the spectrum is very close to the spectrum of the light cone quantisation of the
bosonic string theory
\cite{GGRT_Arvis},
which is only consistent in $D=26$ and $D=3$, and which we shall
refer to as the Nambu-Goto spectrum. (We restrict ourselves
here  to flux tubes that wind around one of the spatial tori, so that there is no Coulomb
interaction, and no extra boundary terms to the world sheet action.) The remarkable
simplicity obtained in lattice calculations (see 
\cite{AABBMT_d4}
and references therein) is now well understood theoretically. Long flux tubes can be
very accurately described by the established series of universal terms in the
world sheet action   
\cite{OA}
(see also
\cite{ML})
while shorter flux tubes can be well understood from
the near-integrability of the world-sheet theory in this limit 
\cite{SD,SD_axion,SD_int}.
The latter framework provides a powerful way to translate the observed energies
into world sheet S-matrix elements and, where appropriate, into extra fields
and interactions in the world sheet action. In particular one ground state in
\cite{AABBMT_d4}
showed large deviations from the simple Nambu-Goto spectrum, in a way that suggested that
it might consist of a massive pseudoscalar world-sheet particle on the background
flux tube. The analysis in
\cite{SD_axion}
using the formalism of
\cite{SD}
shows quite convincingly that this is indeed the case, and that the mass can
be read off from the excitation energy above the absolute ground state to a
good approximation.
Moreover the natural coupling of this pseudoscalar has a topological
interpretation
\cite{SD,SD_axion},
making it natural to call it the world-sheet `axion'.
Recently
\cite{SD_int}
these authors have pointed out that the $N=\infty$ world-sheet
theory might be integrable, but only if it possesses at least one massless mode
in addition to the usual massless `phonons' associated with the string's
spontaneous breaking of the bulk translation symmetry. They also noted that the
observed decrease in the lattice estimates of the axion mass when one goes from
$SU(3)$ to $SU(5)$
\cite{AABBMT_d4}
raises the very interesting possibility that the axion mass might decrease
to zero as $N\to\infty$ thus providing the extra massless mode needed for integrability.
Locating a place for integrability in the planar limit of $SU(N)$ gauge theories
has the potential to provide some analytic control over these theories and so
is an exciting possibility. This motivates the present paper in which
we perform lattice calculations of the axion mass for
larger $N$, so as to see whether this mass vanishes or not in the planar limit.

\section{Lattice calculation}
\label{section_lattice}

\subsection{lattice setup}
\label{subsection_setup}

The lattice calculations in this paper are essentially a direct continuation of the
lattice calculations in
\cite{AABBMT_d4}.
Here we briefly recall that our lattice field variables are $SU(N)$ matrices, $U_l$,
residing on the links $l$ of a periodic $L_x \times L_y \times L_z \times L_t$ lattice,
with lattice spacing $a$. The Euclidean path integral is 
\begin{equation}
Z=\int {\cal{D}}U \exp\{- \beta S[U]\},
\label{eqn_Z}
\end{equation}
where ${\cal{D}}U$ is the Haar measure and we use the standard plaquette action,
\begin{equation}
\beta S = \beta \sum_p \left\{1-\frac{1}{N} {\text{ReTr}} U_p\right\}  
\quad ; \quad \beta=\frac{2N}{g^2}.
\label{eqn_S}
\end{equation}
Here $U_p$ is the ordered product of link matrices around the plaquette $p$. 
We write  $\beta=2N/g^2$, where $g^2$ becomes 
the continuum coupling when $a\to 0$. Monte Carlo calculations are performed 
using a standard Cabibbo-Marinari heat bath
\cite{CM}
plus over-relaxation algorithm.

\subsection{calculating flux tube energies}
\label{subsection_energies}

We calculate the energy of a flux tube that winds around the periodic space-time
in the $x$ direction and which has length $l=aL_x$. The details of the calculation
are exactly the same as described at some length in
\cite{AABBMT_d4}.
Here it is useful to recall that such a flux tube has the following relevant
quantum numbers: spin, $J$, around the axis; a parity, $P_{\shortparallel}$, arising
from $x\to -x$ reflections supplemented by charge conjugation; the $D=2+1$ parity, 
$P_{\perp}$, arising from $(y,z)\to (-y,z)$. We set the momenta  along and 
transverse to the flux tube to zero in this paper.

As described in detail in
\cite{AABBMT_d4}
our operators $\phi(t)$ are essentially Wilson lines that wind around the $x$-torus
and are summed over $y$ and $z$, as well as the starting point in $x$, so as to have
zero momentum. These loops are decorated with various deviations from the direct path
so as to allow us to produce operators $\phi_q(t)$ with various non-trivial quantum
numbers $q$. We extract ground states from the asymptotic decay of the correlators
of such operators,
$\langle \phi_q^\dagger(t) \phi_q(0) \rangle \propto \exp\{-E_q(l)t\}$ as $t\to\infty$
where $E_q(l)$ is the ground state energy of the flux tube with the quantum numbers
$q$. An important constraint is that the statistical error is roughly independent
of $t$, so the exponentially decreasing `signal' can rapidly disappear into the
`noise'. That is to say we need $aE_q(l)$ to be small for a reliable calculation.

In this paper we shall focus on the quantum numbers $J^{P_{\shortparallel}P_{\perp}} = 0^{++}$
and $J^{P_{\shortparallel}P_{\perp}} = 0^{--}$. From the  0$^{++}$ ground state we
extract the string tension, while from the difference between the two ground state
energies we can extract the `axion' mass using
\begin{equation}
  M_A
  \simeq
  E_{0^{--}}(l) -  E_{0^{++}}(l)
\label{eqn_MA}
\end{equation}
as long as $l$ is not too large. (If $l$ is large, then the  lightest $0^{--}$ state
will be the one with massless phonons rather than with a massive axion 
\cite{AABBMT_d4}.) 
This is, of course, an approximation but in practice it is a rather good one: using
eqn(\ref{eqn_MA}) with the spectra in
\cite{AABBMT_d4}
would give  $M_A/\surd\sigma \sim 1.9-2.0$, while the correct value
\cite{SD,SD_axion}
is  $M_A/\surd\sigma \sim 1.85(3)$.

\subsection{calculational strategy}
\label{subsection_strategy}

Ideally we would wish to repeat our calculation in
\cite{AABBMT_d4}
for a range of values of $N$ that is large enough for us to
have confidence in our $N\to\infty$ extrapolation.
Since the cost of pure gauge calculations increases
roughly as $\propto N^3$, this would require substantially
larger computer resources than employed in
\cite{AABBMT_d4}
where our main calculations were for $SU(3)$, with only some
for $SU(5)$. So we shall follow a more limited strategy here
which entails the risk of extra systematic errors, which we will
need to address in detail later on in this paper.

We will perform calculations for $N\leq 12$ which should be
an adequate range, given that the leading large-$N$ correction
should be $\propto 1/N^2$. However to reduce the computational cost
we will work at a larger lattice spacing than in 
\cite{AABBMT_d4}
since this means that we can use lattices that are smaller in lattice
units. We will of course need to check (in Section~\ref{section_results})
that the ensuing lattice spacing corrections are in fact
negligible. Of course if $a$ is larger, then so is the energy
$aE(l)$ in lattice units, which means it will be harder to
extract the energy reliably from the correlators, as discussed in
Section~\ref{subsection_energies}. To compensate for this we
perform the calculation for small values of $l$ where the value
of $aE(l)$ will be modest for those states where $E(l)$ decreases
with decreasing $l$ when $l$ is small.
This is the case for the absolute ground state but is in general 
not the case for excitations of this ground state because
the `phonon' momenta that usually provide the excitation of the flux tube
increase roughly as $\propto 1/l$ as $l$ decreases. Fortunately, the $0^{--}$
excited state we are interested in has no phonons and is a counterexample to this.
So, as we shall see below, we are able
to calculate the ground state energies of the $0^{++}$ and $0^{--}$
flux tubes with reasonable accuracy. (Although we will, unfortunately, not be able
to calculate the energies of other excited states.) Of course we need to use a value
of $l$ where the extraction of the axion mass from the difference between
the $0^{++}$ and $0^{--}$ flux tube energies is justified, and this we
demonstrate in Section~\ref{section_results}, where we shall also
see that it is enough to perform calculations for a single value
of $l$ rather than a range of values. All these restrictions
on our calculations will allow us to perform a reasonably 
accurate calculation even for this large range of $N$.

In addition to the above caveats there is a more general problem with lattice
calculations at large $N$. This is the rapid loss of ergodicity as $N$
increases in exploring fields with different topological charge $Q$.
On a periodic lattice a change in $Q$ requires a fluctuation that
starts as a zero action gauge singularity around some hypercube, then
under the action of the Monte Carlo update deforms into a dislocation,
then grows into an instanton with a small core and then gradually grows
into an instanton with a typically sized core. With such local Monte Carlo
changes the instanton has to pass through a stage where it is small on
physical length scales albeit not small on lattice scales. However at physically
small scales one can do a semiclassical estimate and, as is well known
\cite{EW_QN},
the probability of small instantons is exponentially suppressed in $N$.
(Albeit with some caveats
\cite{MT_QN}.)
Thus the probability of changing $Q$ in a Monte Carlo is also
suppressed exponentially with $N$. This  has long been known
\cite{BLMT01_N,LDDEV01_N,BLMTUW04_Q}.
The question, then, is whether such a freezing of $Q$ has damaging
implications for the calculations in this paper. This is a
particularly relevant question in our context because our `axion'
arises from a topological interaction in the world-sheet action
which can be induced by a $\theta$ term in the bulk $SU(N)$ gauge
theory
\cite{Nair}.
This is the issue we address in detail in Section~\ref{section_topology}.

\section{Results}
\label{section_results}

We perform calculations in $SU(N)$ lattice gauge theories with
$N=2,3,4,5,6,7,8,10,12$. These calculations are performed at a
lattice spacing that is (nearly) constant in physical units with
$a\surd\sigma \simeq 0.300\pm0.001$, as we see from Table~\ref{table_data}.
We also keep the same the length $l=8a$ of the winding flux tube,
i.e. $l\surd\sigma \simeq 2.40$ in physical units. 

\begin{figure}[htb]
\centerline{ \scalebox{1.25}{\scalebox{1.0}{\input{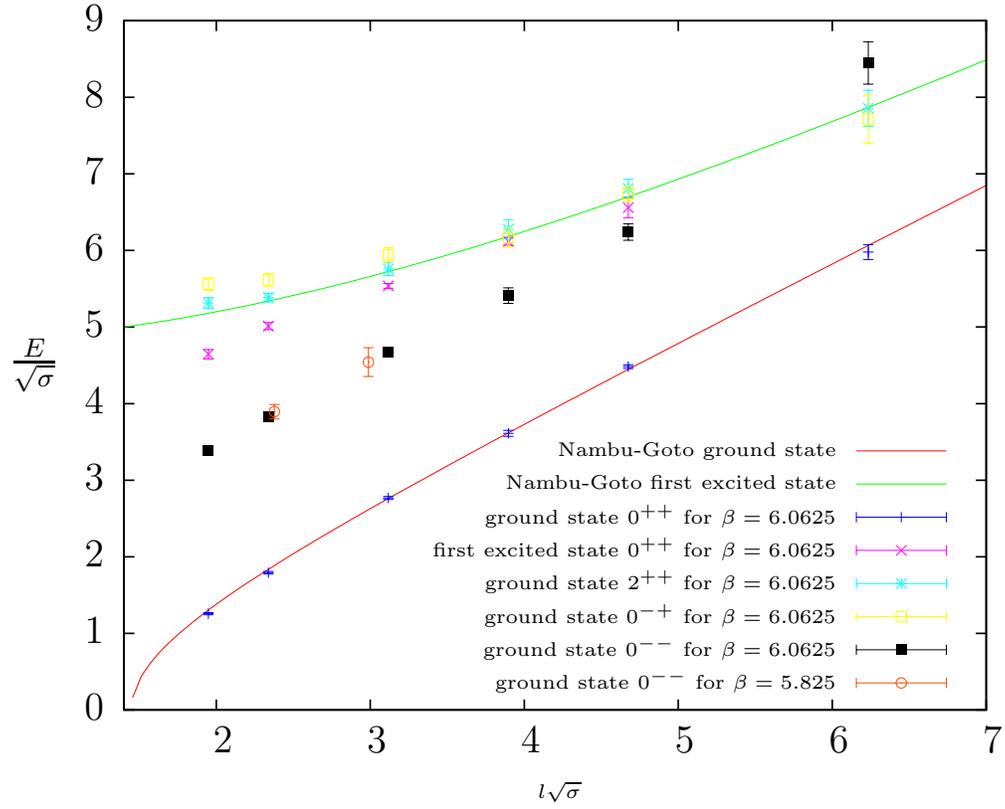}}}}
\caption{Energies of the low-lying flux tube states with the quantum numbers shown, versus
  the length of the flux tube. Results for $SU(3)$ at $\beta=6.0635$ taken from 
  \cite{AABBMT_d4}, compared to our new results for the $J^{P_{\shortparallel}P_{\perp}} = 0^{--}$ flux tube at
  $\beta=5.825$ ({\color{red} $\circ$}).}
\label{fig:su3_comparison}
\end{figure}

We first need to check that using this relatively coarse $a$ does not lead
to significant lattice spacing corrections. To do that we focus on $SU(3)$,
where we have an additional calculation using $l=10a$, and we compare
our results to those we obtained in
\cite{AABBMT_d4}
at a smaller value of $a$, corresponding to $a\surd\sigma \simeq 0.195$ (and
which was itself checked in
\cite{AABBMT_d4}
against results obtained on an even finer lattice with $a\surd\sigma \simeq 0.129$).
The comparison is displayed in Fig~\ref{fig:su3_comparison} and we see that our
values of the $0^{--}$ flux tube energy are perfectly consistent
with those at the smaller lattice spacing. Given that the leading lattice spacing
corrections should be $\propto a^2\sigma$, which decreases by a factor of
$\sim 2.4$ between these two calculations, it appears that any such corrections
must be negligible in our calculation.

The second thing we learn from Fig~\ref{fig:su3_comparison} is that our choice
of $l\surd\sigma \simeq 2.40$ for the calculation of the energy gap
$ \Delta E(l) = E_{0^{--}}(l) - E_{0^{++}}(l)$ is in the range where $\Delta E(l)$
is approximately independent of $l$, and so provides a good estimate of the
`axion' mass using $M_A \simeq  \Delta E(l)$.

We will assume that our above checks for $SU(3)$ carry over to our calculations at
other values of $N$, which is certainly reasonable for $N > 3$. 

\begin{figure}[htb]
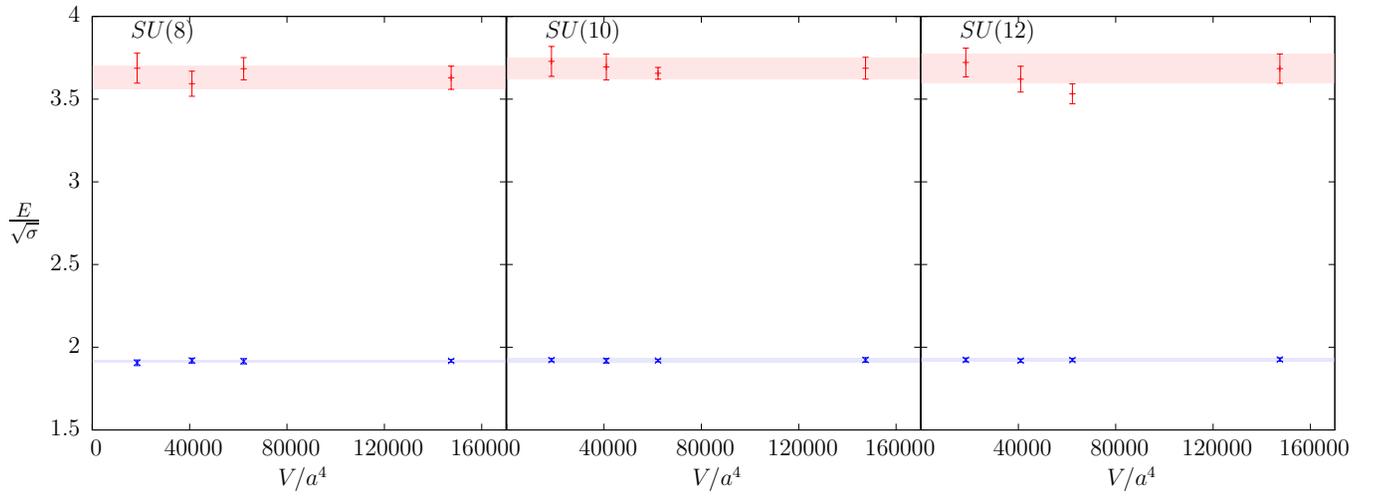

\centerline{ \scalebox{1.25}{\scalebox{0.6}{\input{volume_dependence_SU8.tex}\put(-250,230){$SU(8)$}} \hspace{-3.487cm} \scalebox{0.6}{\input{volume_dependence_SU10.tex}\put(-250,230){$SU(10)$}} \hspace{-3.487cm} \scalebox{0.6}{\input{volume_dependence_SU12.tex}\put(-250,230){$SU(12)$}}}}
\caption{The volume dependence of the energies of the
  $J^{P_{\shortparallel}P_{\perp}} = 0^{++}$  ($\times$, blue) and $0^{--}$ ($+$, red)
  flux tube ground states. The bands correspond to the energy levels obtained with the largest volume i.e.
  $8 \times  24 \times 24 \times 32$.}  
\label{fig:volume_dependence}
\end{figure}

A further check one needs to perform is whether our transverse lattice size is large
enough. To check this we show in Fig.~\ref{fig:volume_dependence} how the
ground state energies of the $0^{++}$ and $0^{--}$ flux tubes vary with the volume,
using the values for $N=8,10,12$ in Table~\ref{table_data} . We see that there is no visible
volume dependence, demonstrating that our initial choice of a $8\times 12\times 12\times 16$
lattice is entirely adequate, at least for these larger values of $N$.

An important final issue that still needs to be addressed has to do with the rapid loss
of topological ergodicity as one increases $N$. This will be dealt with in detail in
Section~\ref{section_topology} where we will argue that this poses no obstacle to our calculations.

\begin{figure}[htb]
\begin	{center}
\leavevmode
\input	{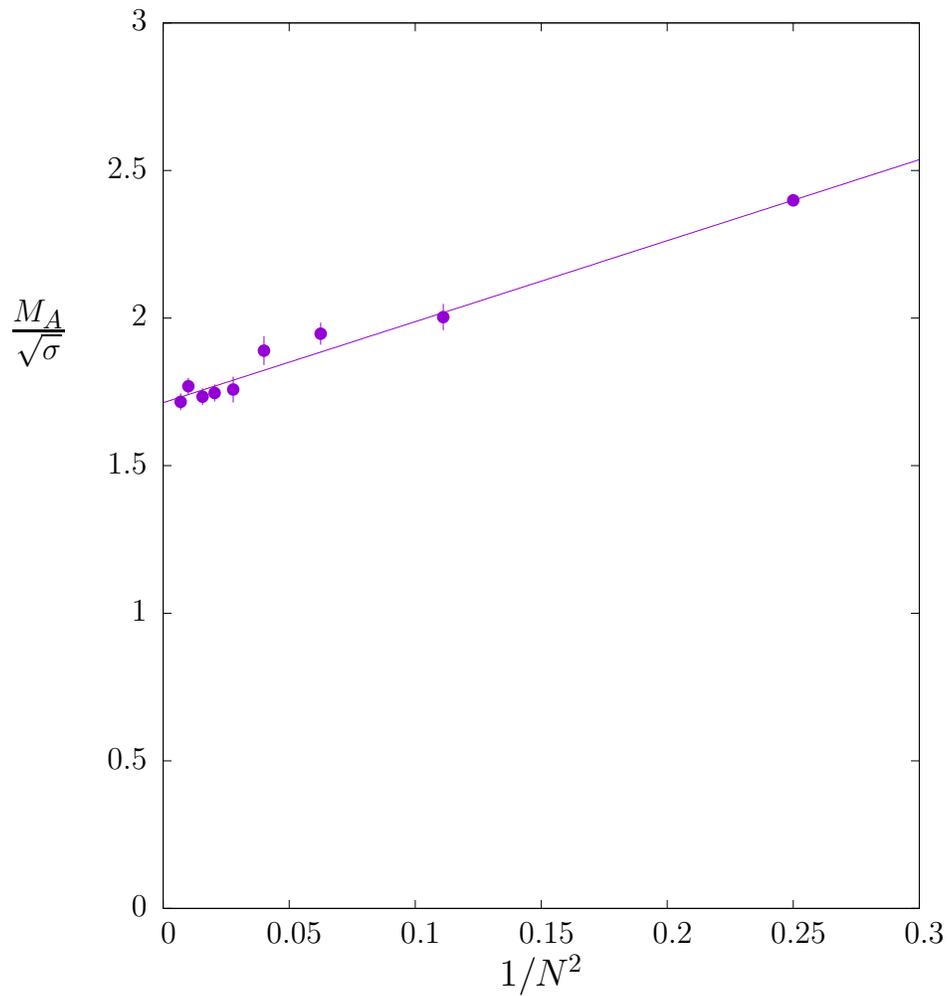}
\end	{center}
\caption{The `axion' mass, using eqn(\ref{eqn_MA}), in units of the string tension, versus
   $1/N^2$ together with a linear extrapolation in $1/N^2$ to $N=\infty$.}
\label{fig:N2_dependence}
\end{figure}

We can  now turn to our estimates of the `axion' mass, $M_A$, using  eqn(\ref{eqn_MA}).
This is shown as a function of $N$ in
Fig.~\ref{fig:N2_dependence}, where we have averaged the values obtained on
the four different volumes for $N=8,10,12$. We fit these values with a linear
function of $1/N^2$, which is the expected leading large-$N$ correction 
\cite{tHooft_N},
giving
\begin{equation}
  \frac{M_A}{\surd\sigma}
  =
  1.713(14) + \frac{2.74(7)}{N^2}
  \qquad ; \quad \chi^2/n_{dof}=1.12 \quad N\in[2,12].
\label{eqn_MA_fitN}
\end{equation}
As shown, the fit has an entirely acceptable $\chi^2$ per degree of freedom, so we
do not need to include any higher order terms in $1/N^2$. (No doubt we would need
to do so if our calculations had much smaller statistical errors.) This fit
is shown in Fig.~\ref{fig:N2_dependence}, and while one sees that ${M_A}/{\surd\sigma}$
does indeed decrease with increasing $N$, it is equally clear
that the `axion' mass does not vanish as $N\to\infty$. 

As a final comment it may be interesting to note that the value of $M_A$ as $N\to\infty$
is about half that of the $N\to\infty$ bulk theory mass gap
\cite{BLMT01_N,BLMTUW04_N,HM_thesis},
although it is somewhat heavier than that at lower $N$.

\section{Topology and non-ergodicity}
\label{section_topology}

As we explained earlier the usual local Monte Carlo algorithms rapidly lose
the ability to change the topological charge $Q$ of a lattice gauge field
as $N$ is increased.
This suppression is related to the suppression of small instantons. If $\rho$
is the size of the instanton and $\lambda(\rho)=g^2(\rho)N$ is the 't Hooft
coupling on the scale of $\rho$, then the density of instantons is
$D(\rho) \propto \exp\{ -8\pi^2 N/\lambda(\rho)\}$ once $\rho$ is small
enough for this semi-classical calculation to be accurate
\cite{EW_QN}.
(There are significant qualifications
\cite{MT_QN}
that we do not enter into here.) Now, a change in $Q$ is accompanied by
an instanton growing from $\rho \sim a$ to $\rho \sim 1/\Lambda$ where
$\Lambda \sim \surd\sigma$ is the physical length scale of the gauge theory.
(Or the reverse process). So this change will be suppressed  by some
factor $\propto \exp\{ -8\pi^2 N/\lambda(\rho)\}$ with $\rho \sim O(a)$.
Since $\lambda(\rho\sim a)$ decreases as $a$ decreases, one way to delay
(in $N$) this suppression is to work at a value of $a$ that is not very small.
In fact this is what we have done in this paper, in part for other reasons.
However this only delays the onset of the problem, and eventually one needs to
confront it.

An important point is that since our theory has a non-zero mass gap,
the value of an observable will only depend on the topological fluctuations
within a finite neighbourhood of that observable, i.e within a distance $\sim 1/\Lambda$.
So the effect on an observable of the total topological charge $Q$ being frozen at
some constant value can be made arbitrarily small by making the space-time volume
sufficiently large. So a direct method to check whether an observable
is affected by this freezing is simply to calculate it on a range of
ever larger volumes to see if it changes. This is in fact the main reason that
we calculated the flux tube energies on four different volumes for our largest three
values of $N$ in Table~\ref{table_data}. The smallest volume of the four is the
`standard' volume we use for $2\leq N\leq 7$. In physical units the space-time
volume orthogonal to the flux tube is already a substantial
$\sim \{3.6/\surd\sigma\}^2\times 4.8/\surd\sigma$. The largest volume is
a much larger $\sim \{7.2/\surd\sigma\}^2\times 9.6/\surd\sigma$.
Nonetheless as we see in Fig.~\ref{fig:volume_dependence} there is no
sign of any change in the relevant flux tube energies as we increase the volume,
strongly suggesting that the topological freezing is already unimportant on our
`standard' volume.

While such direct tests are the most convincing, it is interesting
to see what is actually happening to the topological charge at these larger
values of $N$. Since the method of calculation is fairly standard, we only
briefly summarise it. We recall that the topological charge $Q$ is the integral
over Euclidean space-time of a topological charge density, $Q(x)$, which can be
expressed in terms of the field strengths as
$32 \pi^2 Q(x)=\epsilon_{\mu\nu\rho\sigma} {\mathrm{Tr}}\{F_{\mu\nu}(x) F_{\rho\sigma}(x)\}$.
If on the lattice we replace $F_{ij}(x)$ by the plaquette $U_{ij}(x)$
then we obtain a lattice topological charge density $a^4Q_L(x)$ such
that $Q_L(x) \to Q(x)$ for smooth fields. At finite $\beta$
this lattice measure receives both additive and multiplicative
renormalisation, which can be removed by smoothening the lattice
fields in various ways. We shall employ `cooling'
\cite{MT_cooling}
which consists of performing `Monte Carlo'-type sweeps with the difference
that one locally minimises the action. Under this process $Q$ will
be quasi-stable, since instantons are minima of the continuum action. $Q(x)$ on
the other hand will be gradually deformed as one performs more cooling sweeps.
(For example, neighbouring instantons and anti-instantons can reduce their
action by gradually annihilating each other.) We refer to
\cite{BLMT01_N,BLMTUW04_Q,MT_cooling}
for much more detailed discussions. 

In our calculation we perform 40 cooling sweeps on a sample of 40 or 80 lattice
fields that should be mutually independent for standard physical observables.
From experience we expect that this amount of cooling
will leave the total charge $Q$ unaffected, except for the possible
disappearance of very small instantons. But at larger $N$ there will be almost
none of these. (That is, after all, why $Q$ freezes in the Monte Carlo ensemble.)
Each of these lattice fields
was the last of a separate Monte Carlo sequence used for the calculation of the
flux tube energies, and each started from a near-frozen starting lattice field
with $Q=0$. So if the ergodicity in $Q$ were to be seriously suppressed, then we
would expect to find the fluctuations of $Q$ around $Q=0$ to be suppressed. We list
in Table~\ref{table_dataQ} the values of $\langle Q^2 \rangle$ that we obtain.
We recall that $Q$ is the difference between the number of instantons, $n_+$, and
the number of anti-instantons, $n_-$, and since the correlation length is finite,
and since $\langle n_+ \rangle = \langle n_- \rangle \propto V$, we
will have $\langle Q^2 \rangle \propto V$ once the space-time volume $V$ is large
enough. Indeed in the dilute gas approximation one can readily show that 
$\langle Q^2 \rangle = \langle n_+ \rangle + \langle n_- \rangle = 2\langle n_+ \rangle$.
So it is usual to define the topological susceptibility $\chi_t = \langle Q^2 \rangle /V$,
and to express it in physical units, e.g. $\chi_t^{\frac{1}{4}}/\surd\sigma$. If we
take our largest volume in Table~\ref{table_dataQ} for $SU(8)$ we find
$\chi_t^{\frac{1}{4}}/\surd\sigma \simeq 0.389(14)$ which is entirely consistent 
with the values at smaller $N$ listed in Table 11 of 
\cite{BLMT01_N}
and Table 2 of
\cite{BLMTUW04_Q}.
That is to say, there is no visible suppression in the fluctuations of $Q$ implying
that we still have adequate ergodicity in $Q$ in $SU(8)$ at $\beta=44.355$.
This is in fact no surprise since we chose this value of $a(\beta)$ for our calculations
in the expectation, based on earlier calculations, that this would indeed be the case 
for $SU(8)$. More interesting is $SU(10)$ and $SU(12)$. Since $a\surd\sigma$ is the
same as for $SU(8)$, we would expect that values of $\langle Q^2 \rangle$
to be very similar for $N=8,10,12$ if there continues to be adequate ergodicity
in $Q$. However what we observe in  Table~\ref{table_dataQ} is a dramatic suppression of
$\langle Q^2 \rangle$ as we increase $N$ to $N=10$ and then $N=12$. Clearly
$Q$ is indeed freezing for our highest values of $N$, despite our rather
coarse value of $a(\beta)$.

As an important aside we remind the reader that while changing $Q$ is necessarily
associated with fields that contain small instantons (as described above) and so
will be suppressed at large $N$, one can produce an instanton anti-instanton pair
as a normal unsuppressed long-distance fluctuation. As one updates the field the
separation between such pairs can  grow so that even if $Q=0$ for the total volume
$V$, we expect that in any sufficiently large subvolume $\tilde{V}$ with
$V \gg \tilde{V}$ the integrated
topological charge $\tilde{Q} = \sum_{x\in\tilde{V}} Q(x)$ will have a restricted
susceptibility $\langle {\tilde{Q}}^2\rangle /\tilde{V}$ that equals the true
susceptibility $\langle Q^2 \rangle/V$, when the latter is calculated with `infinite'
statistics so as to overcome any partial non-ergodicity in $Q$.

So, as remarked earlier, the relevant question here is whether the topological
fluctuations in a subvolume that is large enough to contain the flux tube physics
of interest, are significantly suppressed by the observed freezing of $Q$ on the total
volume for $N = 10,12$. Answering this question directly is certainly possible
but would require calculations that take us beyond the scope of this paper.
However we do perform a step in that direction. This is provided by our calculation of
$Q_{abs} \equiv \sum_x |Q(x)|$, whose average  values are listed in Table~\ref{table_dataQ}.
In the dilute gas approximation one can easily show that
$\langle Q_{abs}\rangle = \langle (n_+ + n_-)\rangle = \langle Q^2 \rangle$.
Of course the cooling will tend to decrease $Q_{abs}$ because of the gradual
annihilation of nearby (anti)instanton pairs, and in any case in reality the `gas' 
is surely not dilute. Nonetheless the approximate equality in Table~\ref{table_dataQ}
between $\langle Q_{abs}\rangle$ and $\langle Q^2 \rangle$ for $SU(8)$, suggests that
this argument has an approximate validity, and that $\langle Q_{abs}\rangle$ does give
us a measure of the local density of topological fluctuations. If we now compare
to the values of $\langle Q_{abs}\rangle$ listed in Table~\ref{table_dataQ} for
$SU(10)$ and $SU(12)$, we see that they are almost exactly the same. We take
this as some evidence that the observed onset of a serious non-ergodicity in $Q$
at our largest values of $N$ will not have a significant impact on the topological
fluctuations in the relevant subvolume $\tilde{V}$ as long as $\tilde{V} \ll V$.
For our largest volume,
$V\simeq \{2.4/\surd\sigma\}\times\{7.2/\surd\sigma\}^2\times 9.6/\surd\sigma$,
this inequality is presumably well satisfied.

\section{Conclusions}
\label{section_conclusions}

In this paper we provided a calculation of the world-sheet `axion' mass in
$D=3+1$ $SU(N)$ gauge theories, using the difference between the energies of the
lightest $J^{P_{\shortparallel}P_{\perp}} = 0^{++}$ and $0^{--}$ flux tubes to estimate that mass.
Although our calculations were at a fixed and rather coarse value of the lattice
spacing, comparisons with earlier $SU(3)$ calculations reassured us that
the estimates are reliable. Our calculations covered a much larger range
of $N$ than before, so as to allow a convincing large-$N$ extrapolation.
This required us to address the known rapid loss of lattice Monte Carlo
ergodicity in the topological charge, when $N$ becomes large. We addressed
this directly by performing our calculations on a range of space-time volumes,
and also by calculating the topological charge density, all of which strongly
suggests that the clearly visible loss of this ergodicity does not have a
significant impact on our results.

Our unambiguous  conclusion is that the world sheet `axion' has a finite non-zero
mass at $N=\infty$, and that its mass is roughly half the bulk theory mass gap. That is to
say, it cannot play the role of the extra massless world-sheet mode that is needed
if the $N=\infty$ world sheet theory is to be integrable
\cite{SD_int}.
It needs to be stressed however that the currently available flux tube calculations
\cite{AABBMT_d4}
are incomplete and do not provide accurate calculations of all quantum numbers.
Such calculations will require a basis of flux tube operators that is more
extensive than that employed in
\cite{AABBMT_d4}
and until such a calculation is completed the possibility of some extra massless
world-sheet modes at $N=\infty$ certainly cannot be excluded.

\section*{Acknowledgements}

We would like to thank Sergei Dubovsky, Raphael Flauger and Victor Gorbenko for encouraging
this investigation and for discussions on various aspects of this project. In addition AA
acknowledges Krzysztof Cichy for discussions on topological aspects of this work.
Furthermore we are indebted to participants at both the recent Flux Tubes conference held
at the Perimeter Institute (in May 2015) and the earlier Confining Flux Tubes and Strings
conference held at ECT, Trento (in July 2010). We are
grateful to these institutions for hosting these very productive meetings. AA has been partially
supported by an internal program of the University of Cyprus under the name of BARYONS.
In addition, AA acknowledges the hospitality of the Cyprus Institute where part of this work
was carried out. The numerical computations were carried out on the computing cluster
in Oxford Theoretical Physics.

\vspace*{3.0cm}

\begin{table}[ht]
\begin{center}
\begin{tabular}{|cc|cc|cc|}
\hline Gauge Group & $\beta$ & $L_x\times L_y\times L_z\times L_t$ & $a \sqrt{\sigma}$  & $a E_{0^{++}}$ & $a E_{0^{--}}$\\
\hline  $SU(2)$  & $2.360$ & $8 \times 12 \times 12 \times 16$  & $0.29944(77)$  & 0.5716(38) &  1.290(14) \\ \hline
$SU(3)$  & $5.825$ & $8 \times 12 \times 12 \times 16$  & $0.29733(67)$  &  0.5613(33) & 1.157(25) \\ 
         &         & $10 \times 12 \times 12 \times 16$  & $0.2987(13)$  &  0.7806(81) & 1.357(52) \\ \hline
$SU(4)$  & $10.70$ & $8 \times 12 \times 12 \times 16$  & $0.30007(75)$  &  0.5747(37) & 1.159(21) \\ \hline
$SU(5)$  & $17.00$ & $8 \times 12 \times 12 \times 16$  & $0.29951(57)$  &  0.5720(28) & 1.138(29) \\ \hline
$SU(6)$  & $24.71$ & $8 \times 12 \times 12 \times 16$  & $0.29809(72)$  &  0.5650(35) & 1.089(26) \\ \hline
$SU(7)$  & $33.825$ & $8 \times 12 \times 12 \times 16$  & $0.29935(63)$  &  0.5712(31) & 1.094(17) \\ \hline
$SU(8)$  & $44.355$ & $8 \times 12 \times 12 \times 16$  & $0.29915(98)$  & 0.5702(48) & 1.103(27) \\ 
  &  & $8 \times 16 \times 16 \times 20$  & $0.30047(83)$  & 0.5767(41) & 1.080(23) \\
  &  & $8 \times 18 \times 18 \times 24$  & $0.30007(89)$  & 0.5747(44) & 1.105(20) \\ 
  &  & $8 \times 24 \times 24 \times 32$  & $0.30031(51)$  & 0.5759(25) & 1.090(21) \\ \hline
$SU(10)$  & $69.617$ & $8 \times 12 \times 12 \times 16$  & $0.30078(55)$  & 0.5782(27) & 1.121(27) \\ 
  & & $8 \times 16 \times 16 \times 20$  & $0.30037(77)$  & 0.5762(38) & 1.110(23) \\
  & & $8 \times 18 \times 18 \times 24$  & $0.30049(43)$  & 0.5768(21) & 1.097(11) \\ 
  & & $8 \times 24 \times 24 \times 32$  & $0.30086(77)$  & 0.5786(38) & 1.109(20) \\ \hline
$SU(12)$  & $100.50$ & $8 \times 12 \times 12 \times 16$  & $0.30090(67)$ & 0.5788(33) & 1.120(26) \\ 
  & & $8 \times 16 \times 16 \times 20$  & $0.30045(59)$  & 0.5766(29) & 1.088(23) \\
  & & $8 \times 18 \times 18 \times 24$  & $0.30082(53)$  & 0.5784(26) & 1.063(18)\\ 
  & & $8 \times 24 \times 24 \times 32$  & $0.30112(59)$  & 0.5799(29) & 1.109(27) \\ \hline
\end{tabular}
\caption{Ground state energies of flux tubes of length $l=aL_x$ with
  $J^{P_{\shortparallel}P_{\perp}} =0^{++}$ and $0^{--}$ quantum numbers,
  and also the string tension. For lattices, groups and couplings shown.}
\label{table_data}
\end{center}
\end{table}

\begin{table}[ht]
\begin{center}
  \begin{tabular}{|c|ccc|}
    \hline
Gauge Group & Lattice & $\langle Q_l^2 \rangle$ & $\langle \sum_x |Q_l(x)| \rangle$ \\ \hline
$SU(8)$  & $8 \times 12 \times 12 \times 16$  & 4.16(96)  & 2.55(13)  \\
  &  $8 \times 16 \times 16 \times 20$  & 4.08(72)   &  5.23(15) \\
  &  $8 \times 18 \times 18 \times 24$  & 9.88(1.74) &  8.48(17) \\
  &  $8 \times 24 \times 24 \times 32$  & 27.4(4.0)  &  20.08(19) \\  \hline
$SU(10)$ & $8 \times 12 \times 12 \times 16$  & 1.09(21)  & 2.22(6)  \\
  & $8 \times 16 \times 16 \times 20$  & 2.61(73)   & 5.43(13)  \\
  & $8 \times 18 \times 18 \times 24$  & 7.86(1.96) & 8.21(17)  \\
  & $8 \times 24 \times 24 \times 32$  & 7.66(1.10) & 19.76(17)  \\ \hline
$SU(12)$ & $8 \times 12 \times 12 \times 16$  & 0.07(4)   & 2.09(6)  \\
  & $8 \times 16 \times 16 \times 20$  & 0.35(11)   & 5.22(12)  \\
  & $8 \times 18 \times 18 \times 24$  & 0.47(13)   & 8.23(10)  \\
  & $8 \times 24 \times 24 \times 32$  & 0.48(10)   & 19.47(15)  \\ \hline
\end{tabular}
\caption{ Fluctuations of the total topological charge, $Q$, and the integral of
  the absolute value of the topological charge density, $|Q(x)|$, on various
  lattice volumes for the $N=8,10,12$.}
\label{table_dataQ}
\end{center}
\end{table}

\clearpage

\end{document}